\documentclass[aps,prl,nobibnotes,twocolumn,superscriptaddress,bibliography]{revtex4-2}

\usepackage{amsfonts}
\usepackage{mathrsfs}
\usepackage{amsmath}
\usepackage{color}
\usepackage{graphicx}
\usepackage{bm}
\usepackage{amssymb}
\usepackage{xspace}
\usepackage{epstopdf}
\usepackage{dcolumn}
\usepackage{longtable}
\usepackage{multirow}
\usepackage{float}
\usepackage{comment}

\usepackage[colorlinks=true, letterpaper=true, pdfstartview=FitV, linkcolor=blue, citecolor=blue, urlcolor=blue]{hyperref}

\makeatother

\begin{document}
\title{Double Dirac Nodal Line Semimetal with Torus Surface State }
\author{Xiao-Ping Li}
\affiliation{Key Lab of Advanced Optoelectronic Quantum Architecture and Measurement (MOE), Beijing Key Lab of Nanophotonics \& Ultrafine Optoelectronic Systems, and School of Physics, Beijing Institute of Technology, Beijing 100081, China}
\author{Botao Fu}
\affiliation{College of Physics and Electronic Engineering, Center for Computational Sciences, Sichuan Normal University, Chengdu, 610068, China}
\author{Da-Shuai Ma}
\affiliation{Key Lab of Advanced Optoelectronic Quantum Architecture and Measurement (MOE), Beijing Key Lab of Nanophotonics \& Ultrafine Optoelectronic Systems, and School of Physics, Beijing Institute of Technology, Beijing 100081, China}
\author{Chaoxi Cui}
\affiliation{Key Lab of Advanced Optoelectronic Quantum Architecture and Measurement (MOE), Beijing Key Lab of Nanophotonics \& Ultrafine Optoelectronic Systems, and School of Physics, Beijing Institute of Technology, Beijing 100081, China}
\author{Zhi-Ming Yu}
\email{zhiming\_yu@bit.edu.cn}
\affiliation{Key Lab of Advanced Optoelectronic Quantum Architecture and Measurement (MOE), Beijing Key Lab of Nanophotonics \& Ultrafine Optoelectronic Systems, and School of Physics, Beijing Institute of Technology, Beijing 100081, China}
\author{Yugui Yao}
\email{ygyao@bit.edu.cn}
\affiliation{Key Lab of Advanced Optoelectronic Quantum Architecture and Measurement (MOE), Beijing Key Lab of Nanophotonics \& Ultrafine Optoelectronic Systems, and School of Physics, Beijing Institute of Technology, Beijing 100081, China}

\begin{abstract}
We propose a class of nodal line semimetals that host an eight-fold degenerate double Dirac nodal line (DDNL) with negligible  spin-orbit coupling. We find only 5 of the 230 space groups host the DDNL. The DDNL  can be considered as a combination of two Dirac nodal lines, and  has a trivial Berry phase. This leads to two possible but completely different surface states, namely, a torus surface state covering the whole surface Brillouin zone and no surface state at all.  Based on first-principles calculations, we predict that the hydrogen storage material LiBH is an ideal DDNL semimetal, where  the line resides at Fermi level, is relatively flat in energy, and exhibits a large linear energy range. Interestingly, both the two novel surface states of DDNL can be realized in  LiBH.  Further, we predict that with a magnetic field parallel to DDNL, the Landau levels of DDNL are  doubly degenerate due to Kramers-like degeneracy and have a doubly degenerate zero-mode.
\end{abstract}
\maketitle

\textit{\textcolor{blue}{Introduction}}\textit{.} The past decade has witnessed a tremendous advance in the understanding of topological
band theory, for which one of the most representative and experimentally relevant realization may be topological semimetals, where novel quasi-particles, such as Weyl and Dirac fermion, appear as low-energy excitations around nontrivial crossings formed by conduction and valence bands \cite{Chiu2016Classification-RoMP,toposemi2016,bernevig2018recent,weylDirac}. Various fascinating phenomena associated with  topological semimetals are predicted, such as topologically protected surface states \cite{PhysRevB.83.205101,weylDirac}, unusual optical and magnetic responses \cite{burkov2015chiral,de2017quantized,yu2016predicted,nagaosa2020transport}, density fluctuation plasmons \cite{PhysRevB.75.205418,PhysRevB.91.035114}, and quantized circulation of anomalous shift \cite{PhysRevLett.125.076801}. %

In three-dimensions, the band crossing, in addition to zero-dimensional
(0D) nodal points \cite{murakami2007phase,young2012dirac,wang2012dirac,wang2013three,weng2015weyl,meng2020ternary,bradlyn2016beyond}, also can be 1D nodal line (NL) \cite{burkov2011topological,kim2015dirac,weng2015topological,fang2016topological,13li2016dirac} or even 2D nodal surface \cite{liang2016node,zhong2016towards,wu2018nodal,zhang2018nodal},
protected by corresponding space group (SG) symmetries. The current study
of nodal line semimetals mainly focus on the case that the line is
generated by band inversion, topologically protected by $\pi$ Berry
phase and characterized by drumhead-like surface state. Various topological
nodal line semimetals belonging to this paradigm are predicted, and
some of them have been experimentally confirmed \cite{9ezawa2016loop,huang2016topological,12schoop2016dirac,bian2016topological,16li2017type,xu2017topological,ma2018mirror,he2020ferromagnetic,PhysRevLett.124.016402,jin2020fully}. With multiple symmetries, the nodal line can  take many different  forms in Brillouin zone (BZ), such as higher-order
nodal line, nodal chain, crossed nodal line, nodal box and Hopf-link loop \cite{PhysRevB.99.121106,bzduvsek2016nodal,wang2017hourglass,yu2017nodal,yu2015topological,14kobayashi2017crossing,sheng2017d,fu2018hourglasslike,chen2017topological,yan2017nodal,chang2017topological,chang2017weyl,23yan2017nodal}.

In this work, we theoretically propose another possibility of the nodal line, namely, double Dirac nodal line (DDNL) in 3D systems with negligible spin-orbit coupling (SOC) effect (such as the materials with  atoms of  carbon or even lighter than carbon). This nodal line is four-fold degenerate (eight-fold degenerate if including spin degree of freedom) and resides at certain high-symmetry line in BZ.
For each 2D plane transverse to DDNL (setting as $k_{x}$-$k_{y}$ plane), the band crossing on the line can be considered as a sum of two Dirac point and its effective Hamiltonian may be written as
\begin{eqnarray}
{\cal H} & = & \left[\begin{array}{cc}
h_{D} & h^{\prime}\\
h^{\prime\dagger} & h_{D}
\end{array}\right],\label{eq:hamgen}
\end{eqnarray}
where each entry is a $2\times2$ submatrix and
\begin{eqnarray}
h_{D} & = & v_{x}k_{x}\sigma_{x}+v_{y}k_{y}\sigma_{y},
\end{eqnarray}
with  $\sigma_{i}$ ($i=x,y,z$)  the Pauli matrix, and $v_{x(y)}$  the Fermi velocity in $x(y)$ direction. The two diagonal blocks ($h_{D}$) describe two 2D Dirac points with same topological charge (Berry phase) of $\pi$ in the $k_{x}$-$k_{y}$ plane, and the off-diagonal term $h^{\prime}$ denotes the coupling between the two Dirac points. While Dirac nodal line has been well studied in previous works \cite{kim2015dirac,12schoop2016dirac}, the  DDNL has not yet been proposed.

{\renewcommand\arraystretch{1.3}
\begin{table}[b]
\caption{Space groups allowing for DDNL with  negligible SOC effect.
The DDNL is stabilized by the nonsymmorphic operators of systems. Here, $\alpha \in (0,\frac{1}{2})$}\label{tab1}
\begin{ruledtabular} %
\begin{tabular}{lll}
SG No. & BZ & Location \\
		 \hline
		57& ${\rm \Gamma_{o}}$ & RT: $\{\alpha,\frac{1}{2},\frac{1}{2}\}$  \\
		60& ${\rm \Gamma_{o}}$ & RU: $\{\frac{1}{2},\alpha,\frac{1}{2}\}$  \\
		61& ${\rm \Gamma_{o}}$ & RT: $\{\alpha,\frac{1}{2},\frac{1}{2}\}$, RU: $\{\frac{1}{2},\alpha,\frac{1}{2}\}$, RS: $\{\frac{1}{2},\frac{1}{2},\alpha\}$  \\
     62& ${\rm \Gamma_{o}}$ & RS: $\{\frac{1}{2},\frac{1}{2},\alpha\}$  \\
     205& ${\rm \Gamma_{c}}$ & RM: $\{\frac{1}{2},\frac{1}{2},\alpha\}$  \\
\end{tabular}\end{ruledtabular}
\end{table}
}

Generally, the electron bands in the systems without SOC effect is not degenerate. Hence, the DDNL is rather rare and requires strict symmetries for its realization. Indeed, by an exhaustive searching over 230 SGs \cite{Bradley2009Mathematical-Oxford}, we find that only 5 SGs can exhibit the DDNL, which resides along the high-symmetry line located at  the boundary of BZ. The five SGs and the location of DDNL in BZ are given in Table. \ref{tab1}.  Particularly, for all the five SGs, the DDNL is the only possible degeneracy  at the corresponding high-symmetry line(s), indicating that the number of the electronic bands of materials belonging to these SGs must be a multiple of $8$ (including spin degree of freedom). Hence, for the material belonging to these SGs and having $8n+4$ (with $n$ an integer) electrons, it must be a DDNL semimetal enforced by the filling \cite{watanabe2016filling}.

\begin{figure}[t]
\includegraphics[width=8.2cm]{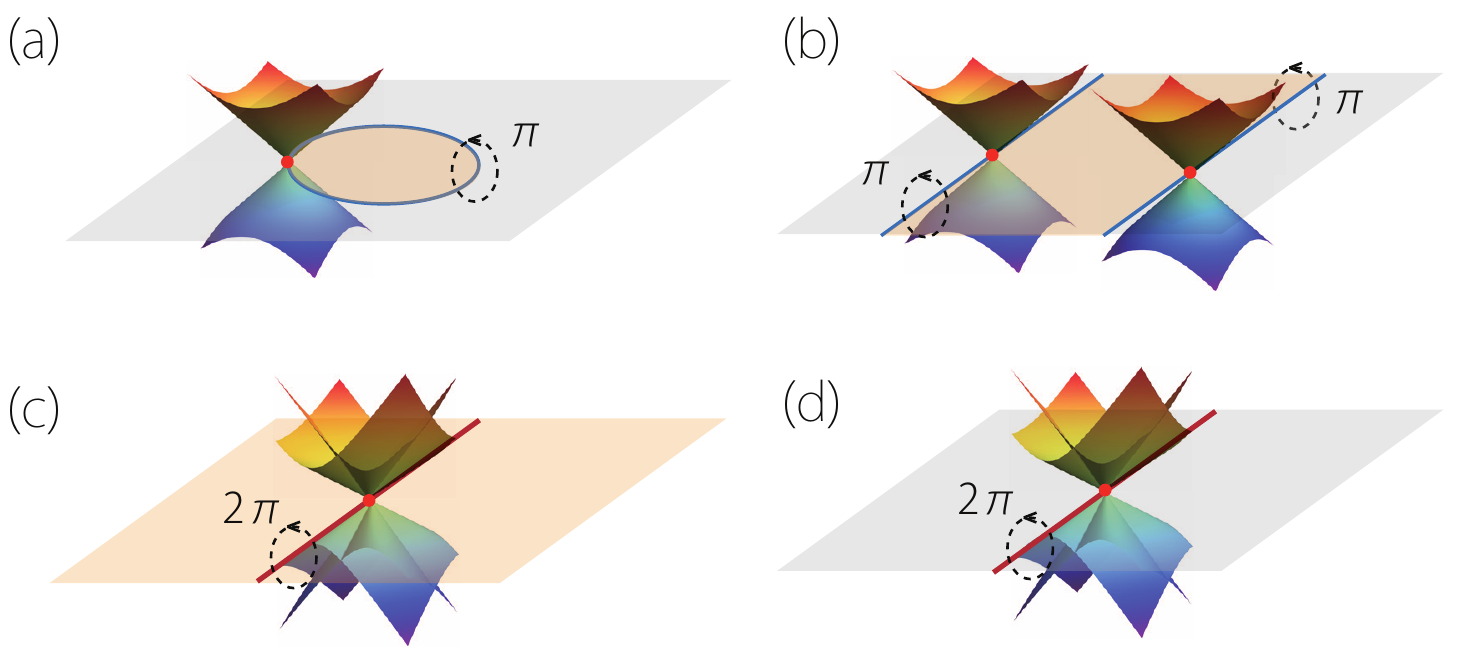} \caption{Schematic showing the surface state of nodal line semimetal, as well as the electronic band around line.
The orange region denotes surface state. (a) Usual Dirac NL with drumhead-like surface state. (b) A Dirac NL traversing bulk BZ should come in pair
and can be considered as a variation of usual Dirac NL due to strong anisotropy.
(c-d) DDNL has two possible but distinct surface states: (c) a torus surface state spanning the whole surface BZ, and  (d) no surface state. \label{fig1}}
\end{figure}

The DDNL is topologically distinct from the usual Dirac nodal line (with  $\pi$ Berry phase) in that its Berry phase is $2\pi$, as it can be considered as a combination of two Dirac nodal lines.  Since the Berry phase is defined mod of $2\pi$, it is trivial for  DDNL. As a consequence,  the DDNL features two distinct states at the boundary  of system, as schematically shown in Fig. \ref{fig1}(c-d), which both are completely different from the drumhead-like surface state in usual NL {[}see Fig. \ref{fig1}(a-b){]}. One is the novel surface state spanning over the whole surface BZ {[}see Fig. \ref{fig1}(c){]}. Since the 2D surface BZ is a torus, such novel surface state then is termed as torus surface state \cite{PhysRevB.99.121106,wang-APS2019}. The other case is that no surface state appears on the surface, even though there exists a NL (e.g. DDNL) in bulk {[}see Fig. \ref{fig1}(d){]}. These two cases are topologically allowed and consistent with the trivial Berry phase of DDNL.
Moreover, we predict that by applying  a magnetic field parallel to the line, the Landau levels of DDNL are  doubly degenerate due to Kramers-like degeneracy and have a doubly degenerate zero-mode, which  shall suggest  pronounced signature in magneto-transport.

We demonstrate our ideas by  the first-principles calculations and model analysis of a concrete material. We find the hydrogen storage material LiBH is an ideal DDNL semimetal candidate, where the DDNL  resides at Fermi level, is relatively flat in energy, and exhibits a large linear energy range. Interestingly, both torus surface state and no surface state simultaneously appear in LiBH material. More importantly, we calculate the LL spectrum of a lattice model based on LiBH and find the doubly degenerate zero-mode LL can be clearly observed. These results indicate that the novel properties of DDNL predicted here should be experimentally detected in LiBH.
Thus, our work not only predicts a new semimetal phase, but also shows an ideal material platform for exploring interesting fundamental physics connected to it.

\begin{figure}[t]
\includegraphics[width=8.2cm]{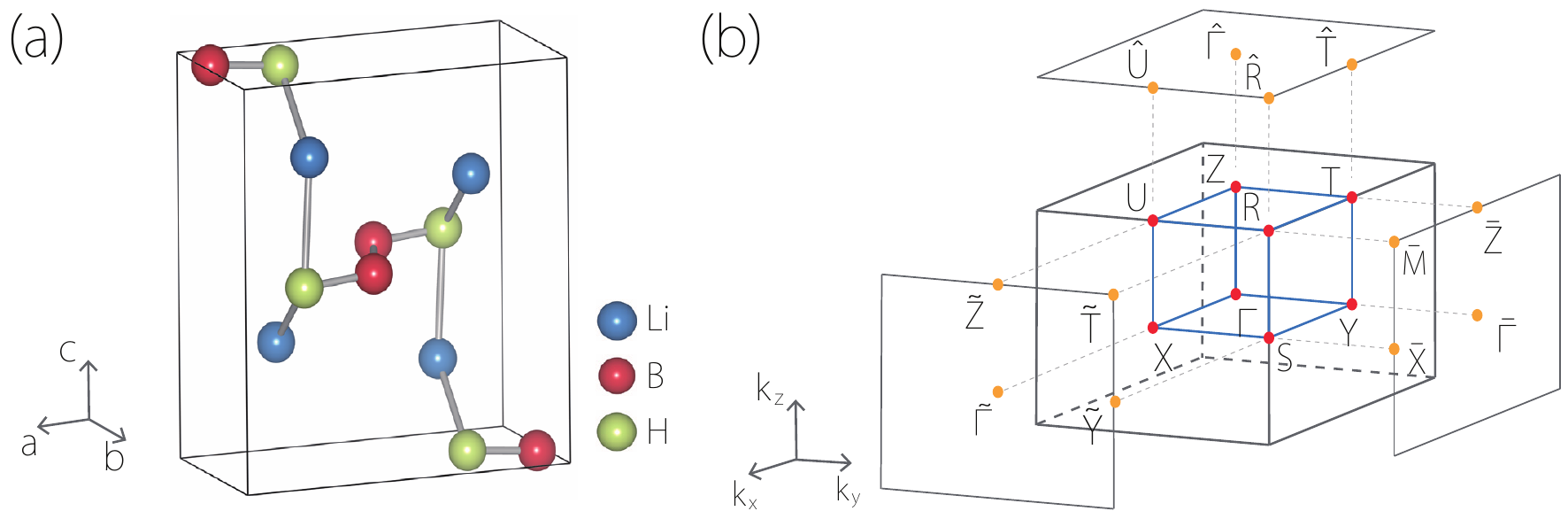} \caption{(a) Crystal structure of LiBH material. (b) Bulk BZ and (100), (010) and (001) surface BZ.
\label{fig2}}
\end{figure}

\textit{\textcolor{blue}{Crystalline structure and electronic bands}}\textit{.}
We motivate our investigation by considering  the hydrogen storage material  LiBH \cite{kang2005candidate}. This material has a orthorhombic structure with SG Pnma (No. $62$), which  is one candidate in Table \ref{tab1}. The primitive cell of LiBH contains in total 12 atoms with Li, B and H residing at the $4c$ Wyckoff positions, as shown in Fig. \ref{fig2}(a). Since all the three atoms: Li, B and H are ${\it lighter}$ than carbon, the SOC effect in LiBH is negligibly small, and we virtually obtain a spinless system. This is a precondition for realizing DDNL, as the DDNL is not robust against SOC \cite{Bradley2009Mathematical-Oxford}.
Moreover, the electron number of LiBH is $20$ ($=2 \times 8+4$). Therefore, the LiBH material is a filling-enforced  DDNL semimetal with  the line appearing around Fermi level.
The lattice constants obtained from our first-principles calculations are $a=6.2 \ \mathring{\mathrm{A}}$,
$b=3.0\  \mathring{\mathrm{A}}$ and $c=6.3 \ \mathring{\mathrm{A}}$ \cite{sm}, consistent with previous result \cite{kang2005candidate}.
The symmetry operators of SG $62$ are generated by two screw rotations $\widetilde{{\cal C}}_{2z}=\left\{ C_{2z}|\frac{1}{2}0\frac{1}{2}\right\} $ and $\widetilde{{\cal C}}_{2y}=\left\{ C_{2y}|0\frac{1}{2}0\right\} $,
and a spatial inversion ${\cal P}$.
The material also has time-reversal symmetry ${\cal T}$ with ${\cal T}^{2}=1$, corresponding to the negligible  SOC effect in LiBH.

\begin{figure}[t]
\includegraphics[width=8.8cm]{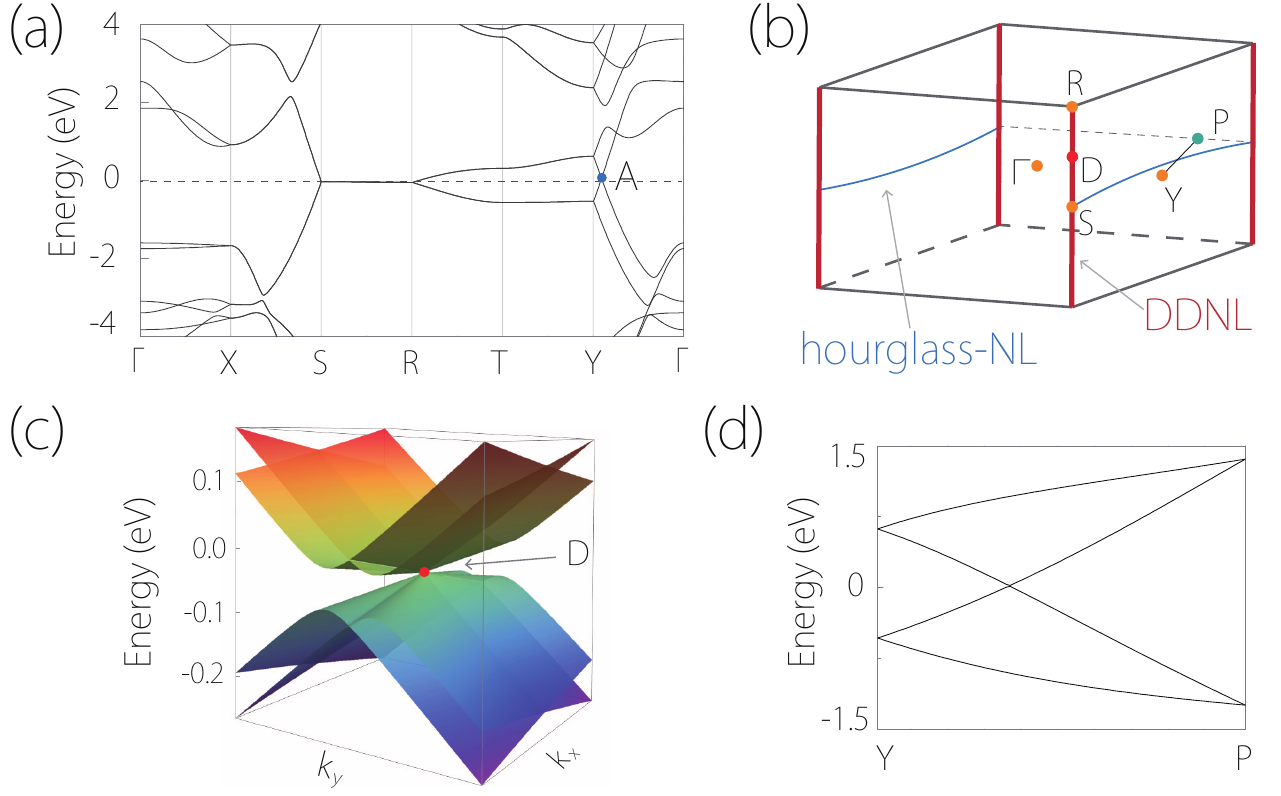} \caption{(a) Electronic band structure of LiBH. (b) Schematic  showing  DDNL and hourglass NL. The electronic band (c) around a generic point (D) on DDNL and (d) a generic path passing  through hourglass NL. \label{fig3}}
\end{figure}

The calculated electronic band structure of LiBH without SOC is presented
in Fig.~\ref{fig3}(a).
It is clearly shown that this material is a DDNL semimetal with  the line  appearing  at RS path, consistent with symmetry  analysis (see Table \ref{tab1}). A remarkable feature of the electronic bands of LiBH is that its low-energy spectra is roughly symmetric about Fermi level,   indicating that LiBH has an approximate chiral symmetry \cite{sheng2019two}.
From Fig.~\ref{fig3}(a), one observes that  there exist two band crossings almost cutting  the Fermi level.
We first consider the linear crossing (labelled as $A$) on  $\Gamma$Y path. Due to the presence of $\widetilde{{\cal M}}_{z}=\widetilde{{\cal C}}_{2z}{\cal P}$
and  ${\cal P}{\cal T}$ symmetry, point $A$ can not appear in isolation but reside on a NL in the $k_{z}=0$ plane {[}see Fig.~\ref{fig3}(b){]}.
As discussed in Ref. \cite{takahashi2017spinless}, this NL is formed by the neck crossing-point of the hourglass-like dispersion {[}see Fig.~\ref{fig3}(d){]} and then is termed as hourglass NL. The hourglass NL can move in the $k_{z}=0$ plane but is unremovable as its two endpoints are pinned at $S$ point.

The second and also the most striking feature of LiBH is the four-fold (eight-fold if including spin) degenerate crossing along RS path, giving rise to a DDNL at Fermi level. The DDNL is rather flat with an energy variation of $20$ meV, due to the approximate chiral symmetry,  and exhibits highly dispersive bands in the plane normal to the line, as shown in  Fig.~\ref{fig3}(a) and~\ref{fig3}(c).
Therefore,   LiBH provides an ideal material platform to explore the novel physics associated with DDNL and
hourglass NL  in experiments. Although the properties of hourglass NL are similar to usual NL \cite{wang2017hourglass,fu2018hourglasslike}, the DDNL  would feature novel phenomena distinguished from usual NL, due to a trivial Berry phase and  completely different low-energy spectrum.
Hence, in the following we  mainly focus on the investigation of the interesting properties of DDNL.

\textit{\textcolor{blue}{DDNL and effective Hamiltonian}}\textit{.}
Since the LiBH material exhibits three orthogonal two-fold screw rotation axes, the electronic bands at all the three boundary planes ($k_{x,y,z}=\pi$ plane) are at least doubly degenerate and the whole bulk BZ is covered by nodal surfaces \cite{wu2018nodal,yu2019circumventing}, as shown in  Fig.~\ref{fig3}(a). The DDNL sits at the hinge between $k_{x}=\pi$ and $k_{y}=\pi$ planes, e.g. the high-symmetry path RS, and then is formed by the crossing of two nodal surfaces. For a generic point ($D$) on RS, its symmetry operators can be generated by $\widetilde{{\cal C}}_{2z}$, $\widetilde{{\cal M}}_{y}=\left\{ {\cal M}_{y}|0\frac{1}{2}0\right\}$ and a combined operator ${\cal A}=\widetilde{{\cal C}}_{2y}{\cal T}$.
The algebra satisfied by the three generators at $D$ point is equivalent to that of the three generators at $S$ point, as both $D$ and $S$ are interior  points of RS line \cite{Bradley2009Mathematical-Oxford}. Then,  for concise we consider the algebra of the three generators at $S$ point to explain   the appearance  of DDNL, which can be written  as
$\widetilde{{\cal C}}_{2z}^{2}=\widetilde{{\cal M}}_{y}^{2}=1$,
${\cal A}^{2}=-1$ and
\begin{eqnarray}
\widetilde{{\cal M}}_{y}\widetilde{{\cal C}}_{2z}=-\widetilde{{\cal C}}_{2z}\widetilde{{\cal M}}_{y},\  & \widetilde{{\cal C}}_{2z}{\cal A}={\cal A}\widetilde{{\cal C}}_{2z},\  & \widetilde{{\cal M}}_{y}{\cal A}=-{\cal A}\widetilde{{\cal M}}_{y}.
\end{eqnarray}
The Bloch states at $S$ point can be chosen as the eigenstates of $\widetilde{{\cal C}}_{2z}$,
denoted as $|c_{2z}\rangle$ with $c_{2z}=\pm1$  the eigenvalue of $\widetilde{{\cal C}}_{2z}$. Since $\widetilde{{\cal C}}_{2z}$ anticommutes with $\widetilde{{\cal M}}_{y}$, the two states $|1\rangle$
and $|-1\rangle$ would be degenerate, as $|\pm1\rangle=\widetilde{{\cal M}}_{y}|\mp1\rangle$.
Also as $\widetilde{{\cal C}}_{2z}$ commutes with ${\cal A}$
and ${\cal A}^{2}=-1$, the state $|\pm1\rangle$ and its Kramers-like
partner ${\cal A}|\pm1\rangle$ are linearly independent. Hence, the
four states $\{|1\rangle,\ |-1\rangle,\ {\cal A}|1\rangle,\ {\cal A}|-1\rangle\}$
must be degenerate at the same energy, forming a DDNL along RS
path. Based on the quartet basis, the matrix representations of the
generators can be expressed as%
\begin{eqnarray}
\widetilde{{\cal C}}_{2z}=\sigma_{0}\otimes\sigma_{z},\ \  & \widetilde{{\cal M}}_{y}=\sigma_{z}\otimes\sigma_{x},\ \  & {\cal A}=i\sigma_{y}\otimes\sigma_{0}{\cal K},
\end{eqnarray}
with ${\cal K}$ the complex conjugation. With the standard approach   \cite{bradlyn2016beyond,wu2020higher},
the effective $k\cdot p$ Hamiltonian 
for a generic point on DDNL  can be obtained as
\begin{eqnarray}
{\cal H}_{\text{DDNL}} & = & (c_{1}+c_{2}k_{z})+\left[\begin{array}{cc}
h_{D} & h^{\prime}\\
h^{\prime\dagger} & h_{D}
\end{array}\right],\label{eq:hamDDNL}
\end{eqnarray}
with $h_{D}=c_{3}k_{x}\sigma_{x}+c_{4}k_{y}\sigma_{y}$ and $h^{\prime}=\alpha k_{x}\sigma_{y}+\beta k_{y}\sigma_{x}$.
Here, $c_{i=1,2,3,4}$ is real parameter, and $\alpha$ and $\beta$
are complex parameters. Clearly, the obtained Hamiltonian (\ref{eq:hamDDNL})
is consistent with Eq. (\ref{eq:hamgen}) in Introduction, which directly demonstrates the existence of DDNL in LiBH. Due to nonvanishing
off-diagonal term $h^{\prime}$, the two Dirac cones described by $h_D$ would  split at
a general momentum point, while are sticked together along $k_{x(y)}=0$
axis [see Fig.~\ref{fig3}(c)], resulting from the nodal surfaces
on boundaries.

\begin{figure}[t]
\includegraphics[width=8.8cm]{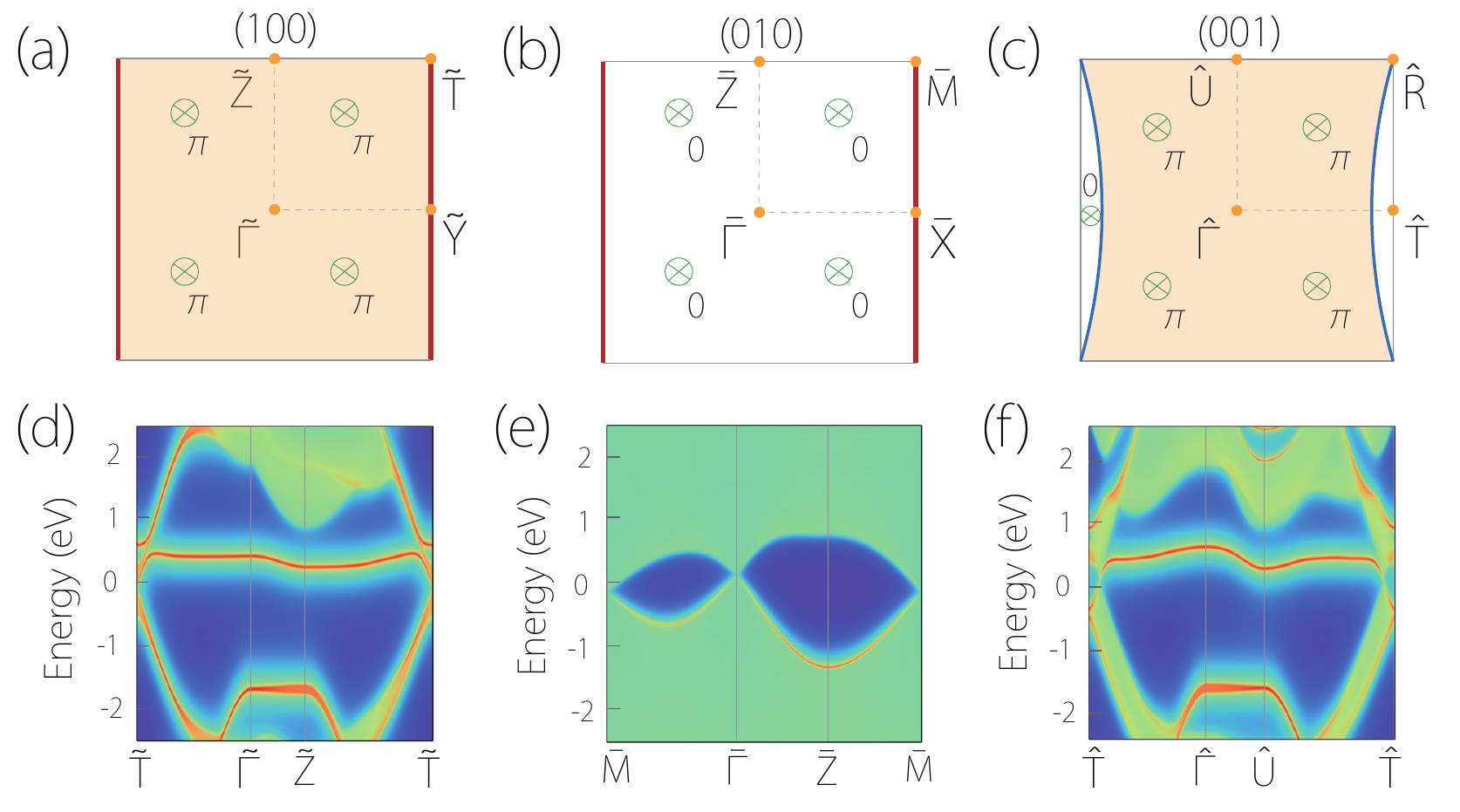} \caption{(a-c) Schematic figures of the surface state for (100), (010) and (001) surfaces. The values 0 and $\pm \pi$ in (a-c) are  the Zak phase for lines normal to the corresponding surface.
(d-f) Surface spectra on (100), (010) and (001) surfaces. In (d), the surface state spans the whole surface BZ, leading to the torus surface state, while in (e) no surface state can be observed.
\label{fig4}}
\end{figure}

\textit{\textcolor{blue}{Torus surface state}}\textit{. } We then explore
the surface state of LiBH. It is has been extensively shown that NL semimetals feature drumhead-like surface state at sample boundary, due to  $\pi$ Berry phase of the line. This is the case for the hourglass NL in LiBH, which has $\pi$ Berry phase and leads a drumhead-like surface state at $(001)$ surface, as shown in Fig.~\ref{fig4}(c,f).
However, both $(100)$ and $(010)$ surfaces, which are parallel to DDNL, do not exhibit drumhead-like surface state.
More surprisingly, the $(100)$ surface has a torus surface state covering the whole $(100)$ surface BZ, as shown in Fig.~\ref{fig4}(a,d), and in sharp contrast, the $(010)$ surface does not have any surface
state [Fig.~\ref{fig4}(b,e)].

The surface state in NL semimetal generally is protected by quantized $\pi$ Zak phase, which is the Berry phase of a straight line normal to the surface and crossing the bulk BZ \cite{chan20163}. Here, due to the presence of ${\cal P}$ symmetry in LiBH, the Zak phase is quantized to be $0$ or $\pi$, corresponding two topologically distinct phases.

We first study the surface state on $(100)$ surface. Starting from
the Zak phase ${\cal Z}(k_{y}=0,k_{z})$ (with $k_{z}\neq0$) of a
straight line transverse to $(100)$ surface, which is obtained as
$\pi$ in LiBH material [see Fig. \ref{fig4}(a)]. By moving the straight line along $k_{y}$-axis,
the Zak phase shall not change its value until the line passes through
DDNL at RS path ($k_y=\pi$). The changed value equals to the Berry phase of
DDNL, namely, $2\pi$. This means that one always has ${\cal Z}(k_{y}>\pi,k_{z})={\cal Z}(k_{y}<\pi,k_{z})$,
as the Zak phase is defined mod $2\pi$. Similarly, by moving the
straight line along $k_{z}$-axis, the Zak phase would change its
value by zero when the line crosses over the hourglass NL at $k_{z}=0$ plane. Therefore, the Zak phase ${\cal Z}(k_{y},k_{z})=\pi$ for any gapped state, resulting in a torus surface state in $(100)$ surface, as shown in Fig.~\ref{fig4}(a,d). Similar analysis applies for $(010)$ surface, except that ${\cal Z}(k_{x},k_{z})=0$ for each gapped state. Hence, no surface state can be found on this surface, as shown in Fig.~\ref{fig4}(b,e). These unusual surface properties also can be inferred from the geometry structure of DDNL. Unlike the usual NL shown in Fig. \ref{fig1}(a-b) and Fig. \ref{fig4}(c) , the projection of DDNL can not separates the corresponding surface BZ as two parts {[}see Fig.~\ref{fig4}(a,b){]}, as the BZ is periodic. Hence, for the surface parallel to DDNL, the whole surface BZ would share same topological properties, leading to a torus surface state or no surface state on the boundary of system.

\begin{figure}[t]
\includegraphics[width=8.8cm]{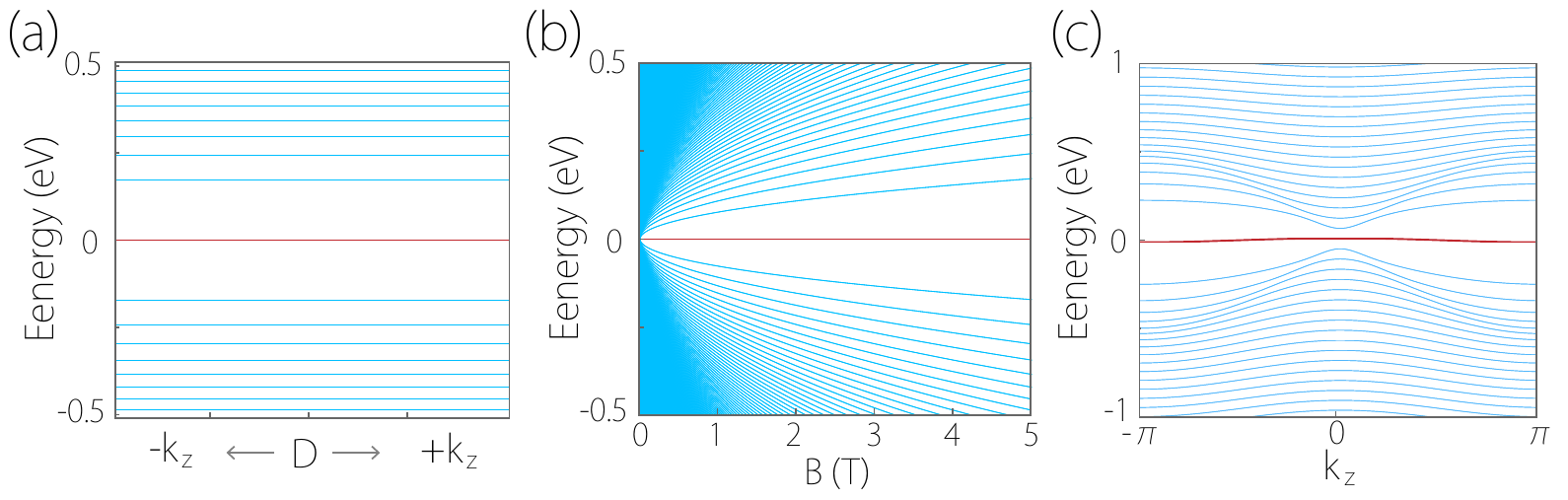} \caption{(a-b) Landau spectrum with (a) different $k_z$ ($B=5$ T) and (b) different $B$ (at $k_z=0$ plane) based on the effective model (\ref{eq:hamDDNL}). (c) Landau spectrum of a lattice model based on LiBH with $B=0.01\frac{\phi_{0}}{a\times b}$($\phi_{0}$ is magnetic flux quantum $h/e$). 
The red curves denote doubly degenerate zero-mode LLs. The calculation details are presented in \cite{sm}.   \label{fig5}}
\end{figure}

\textit{\textcolor{blue}{Landau spectrum}}\textit{. }
The topological feature of NL also can reflect in its magnetotransport \cite{yang2018quantum,zhang2018hybrid}. By applying a magnetic field parallel to DDNL, the electronic bands are  quantized into discrete Landau levels (LLs). 
Generally, one can investigate the LLs of DDNL in a $k_z$-fixed plane, where the low-energy Hamiltonian is captured by Eq. (\ref{eq:hamDDNL}). It is well known that the Dirac Hamiltonian $h_D$ features a zero-mode LL \cite{Goerbig_RMP}. One may wonder that the zero-mode LL would not appear in DDNL, due to the presence of the coupling $h^{\prime}$ between the two Dirac equation. However, this is not the case. We find that the DDNL also has a zero-mode LL for each $k_z$-fixed plane \cite{sm}, as shown in Fig. \ref{fig5}(a,b). Moreover, these zero-mode LLs are  doubly degenerate. While the  $B$  field   breaks both ${\cal{T}}$ and ${\widetilde{{\cal C}}_{2y}}$ symmetries, it preserves ${\cal{A}}$ ($={\widetilde{{\cal C}}_{2y}}{\cal T}$) symmetry, and  then all the LL bands (including zero-mode LLs) of DDNL are doubly degenerate resulting from the  Kramers-like degeneracy produced by ${\cal{A}}^2=-1$.
Since the degeneracy of the two zero-mode LLs  is protected  by ${\cal{A}}$ symmetry, it would be  robust against $B$  field and always exist regardless of the filed strength.
Given the fact that the zero-mode LL in graphene  gives rise to many novel phenomena \cite{Goerbig_RMP}, one can expect the DDNL may exhibit interesting  magnetoresponses distinct from the usual Dirac NLs and also 2D Dirac semimetal.

Particularly, the low-energy spectrum of LiBH is so clean that the LL properties  proposed here shall  be  observed in it.  We further demonstrate it by calculating the LL spectrum of a lattice model based on LiBH material. The calculated results are given  in Fig. \ref{fig5}(c), where a doubly degenerate LL with almost flat dispersion occurring at the Fermi level, corresponding to the doubly degenerate zero-mode LL.
This strongly suggests that the unusual LL spectrum of DDNL  can be detected  in LiBH by magnetotransport measurements.

\textit{\textcolor{blue}{Conclusion}}\textit{. }In summary, we propose a new class of semimetal phase: DDNL in LiBH material. The DDNL can be considered as a combination of two Dirac NLs and exhibits many distinct phenomena, such as torus surface state and unusual LL spectrum. In particular, we predict LiBH material is an ideal DDNL semimetal and demonstrate   that the novel phenomena of DDNL predicted here  can be clearly observed in LiBH material.
Moreover, as the torus surface in LiBH is rather flat in energy {[}see Fig.~\ref{fig4}(a){]}, it will be interesting to explore possible unconventional superconductivity, correlation effect and magnetism in LiBH (100) surface.

\acknowledgements
The authors thank J. Xun for helpful discussions.  This work is supported by the National Key R\&D Program of China (Grant Nos. 2020YFA0308800, 2016YFA0300600), the NSF of China (Grants Nos. 12061131002, 11734003), the Strategic Priority Research Program of Chinese Academy of Sciences (Grant No. XDB30000000) and Beijing Institute of Technology Research Fund Program for Young Scholars.
\bibliography{ref}

\end{document}